\documentclass[12pt]{article}
\usepackage{epsf,a4}
 
\newcommand{\half}{\mbox{\small $\frac{1}{2}$}}    % half
\newcommand{\third}{\mbox{\small $\frac{1}{3}$}}   % third
\newcommand{\quarter}{\mbox{\small $\frac{1}{4}$}} % quarter
\newcommand{\msbar}{\mbox{\tiny $\overline{MS}$}}  % Msbar
\newcommand{\mom}{\mbox{\tiny $MOM$}}              % MOM
\def\lsim{\mathrel{\rlap{\lower4pt\hbox{\hskip1pt$\sim$}}
    \raise1pt\hbox{$<$}}}                % less than or approx. symbol
\def\gsim{\mathrel{\rlap{\lower4pt\hbox{\hskip1pt$\sim$}}
    \raise1pt\hbox{$>$}}}                % greater than or approx. symbol
\begin{document}

\title{
\vspace{-2.5cm}
\flushleft{\normalsize DESY 99-097} \\
\vspace{-0.35cm}
{\normalsize HLRZ 99-30} \\
\vspace{-0.35cm}
{\normalsize HUB-EP-99/31} \\
\vspace{-0.35cm}
{\normalsize FUB-HEP/3-99} \\
\vspace{-0.35cm}
{\normalsize TPR-99-09} \\
\vspace{-0.35cm}
{\normalsize August 1999} \\
\vspace{0.5cm}
\centering{\Large \bf A Lattice Determination of Light Quark Masses}}
%\\[1em]}

\author{\large M.~G\"ockeler$^1$, R.~Horsley$^2$, H.~Oelrich$^3$,
                                                  D.~Petters$^{3,4}$, \\
               D.~Pleiter$^{3,4}$, P.~E.~L.~Rakow$^1$, G.~Schierholz$^{3,5}$
               and P.~Stephenson$^6$\\[2em]
        \small
        $^1$ Institut f\"ur Theoretische Physik, Universit\"at Regensburg,\\
        \small
                    D-93040 Regensburg, Germany\\[0.5em]
        \small
        $^2$ Institut f\"ur Physik, Humboldt-Universit\"at zu Berlin,\\
        \small
                    D-10115 Berlin, Germany\\[0.5em]
        \small
        $^3$ Deutsches Elektronen-Synchrotron DESY,\\ 
        \small
                    John von Neumann-Institut f\"ur Computing NIC, \\
        \small
                    D-15735 Zeuthen, Germany\\[0.5em]
        \small
        $^4$ Institut f\"ur Theoretische Physik,
        \small
                    Freie Universit\"at Berlin,\\
        \small
                    D-14195 Berlin, Germany\\[0.5em]
        \small
        $^5$ Deutsches Elektronen-Synchrotron DESY,\\ 
        \small
                    D-22603 Hamburg, Germany\\[0.5em]
        \small
        $^6$ Dipartimento di Fisica,
                    Universit\`a degli Studi di Pisa \& INFN,\\
        \small
                    Sezione di Pisa, I-56100 Pisa, Italy }
         
%\date{\today}
\date{}

\maketitle

\begin{abstract}
A fully non-perturbative lattice determination of the up/down
and strange quark masses is given for quenched  QCD using both,
$O(a)$ improved Wilson fermions and ordinary Wilson fermions. 
For the strange quark mass with $O(a)$ improved fermions we obtain
$m^{\msbar}_s(\mu=2\,\mbox{GeV}) = 105(4)\,\mbox{MeV}$,
using the interquark force scale $r_0$. Due to quenching
problems fits are only possible for quark masses larger than the 
strange quark mass. If we extrapolate our fits to the up/down quark 
mass we find for the average mass 
$m^{\msbar}_l(\mu=2\,\mbox{GeV}) = 4.4(2)\,\mbox{MeV}$.
\end{abstract}

\clearpage

% ----------------------------------------------------------------------

\section{Introduction}

Some of the least known parameters in the Standard Model are the
light quark masses $m_u$, $m_d$ and $m_s$. Their 
phenomenological values have been discussed since the early days
of the quark model. Paradoxically, the values of the later discovered
heavier quarks are more accurately known \cite{caso98a}.
The reason is that the connection between light quark masses
and observables is highly non-perturbative. This means that
the lattice approach is an appropriate technique for this problem.

In this paper we shall present a completely non-perturbative
determination of light quark masses. The recent major step forward has
been the non-perturbative lattice determination of
the renormalisation constants of the mass operators.
Also, due to the increase in available computer time,
a more reliable continuum extrapolation is now possible.

This paper is organised as follows. In section~\ref{define_qm} we
discuss the definition of the quark mass and its renormalisation
group behaviour. Transcribing lattice data to physical units
requires a scale to be set. For quenched QCD this problem
is discussed in section~\ref{scale_digression}.
The lattice technique for obtaining the quark masses and
their renormalisation is presented in section~\ref{qm_determination}.
In section~\ref{numerical_results} we give our results,
and in section~\ref{results} we extrapolate them to the continuum
limit to remove residual discretisation effects.
We perform the calculations for both, $O(a)$ improved
fermions and for Wilson fermions. Both should extrapolate to
the same continuum result, and thus we have a consistency check
between the two methods. We have previously used tadpole improved 
perturbation theory
to compute the renormalisation constants. In section~\ref{TI}
we test the validity of this approach. Finally, in
section~\ref{conclusions} we give our conclusions.

% ----------------------------------------------------------------------

\section{Defining the quark mass}
\label{define_qm}

Due to confinement quarks are not eigenstates of the
QCD Hamiltonian and are thus not directly observable.
A definition of the quark mass
from an experiment thus means prescribing the measurement procedure.
Theoretically this is equivalent to giving a renormalisation scheme
${\cal S}$ and scale $M$.
Conventionally, quark masses are given in a mass independent scheme,
such as the $\overline{MS}$ scheme, at some given scale $\mu$,
commonly taken as $2\,\mbox{GeV}$ \cite{caso98a}.
In a general mass independent scheme ${\cal S}$
the renormalised quark mass is given by
\begin{equation}
   m^{\cal S}(M) = Z_m^{\cal S}(M) m_{bare}.
\end{equation}
The running of this renormalised quark mass as the scale $M$
is changed is controlled by the $\beta$ and $\gamma$ functions
in the renormalisation group equation. These are defined
as scale derivatives of the renormalised coupling and
mass renormalisation constant as
\begin{eqnarray}
   \beta^{\cal S} \left(g^{\cal S}(M) \right) &\equiv&
                 \left. {\partial g^{\cal S}(M) \over
                         \partial \log M }\right|_{bare},
                                   \label{beta_def} \\
   \gamma_m^{\cal S} \left(g^{\cal S}(M) \right) &\equiv&
                 \left. {\partial \log Z_m^{\cal S}(M) \over
                         \partial \log M }\right|_{bare},
                                   \label{gamma_def}
\end{eqnarray}
where the bare parameters are held constant.
These functions are given perturbatively as power series expansions
in the coupling constant. The expansion is now known to four loops
in the $\overline{MS}$ scheme \cite{ritbergen97a,vermaseren97a}.
We have
\begin{eqnarray}
   \beta^{\msbar}(g)  &=& - b_0g^3 - b_1g^5
                          - b_2^{\msbar}g^7 - b_3^{\msbar}g^9 - \ldots,
                                    \nonumber \\
   \gamma_m^{\msbar}(g) &=&   d_{m0}g^2 + d_{m1}^{\msbar}g^4
                          + d_{m2}^{\msbar}g^6 + d_{m3}^{\msbar}g^8 + \ldots,
\end{eqnarray}
where (for quenched QCD)
\begin{eqnarray}
   \begin{array}{lllllll}
      b_0          &=& {11\over (4\pi)^2},
                   & &
      b_1          &=& {102\over (4\pi)^4},   \\
      b_2^{\msbar} &=& { 1\over (4\pi)^6}\left[ {2857\over 2} \right],
                   & &          
      b_3^{\msbar} &=& {1 \over (4\pi)^8}
                             \left[ {149753\over 6} + 3564\zeta_3 \right],
   \end{array}
\end{eqnarray}
and
\begin{eqnarray}
   \begin{array}{lllllll}
      d_{m0}          &=& - {8\over (4\pi)^2},
                      & &
      d_{m1}^{\msbar} &=& - {404\over (4\pi)^4},  \\
      d_{m2}^{\msbar} &=& - {2498\over (4\pi)^6},
                   & &
      d_{m3}^{\msbar} &=& - {1 \over (4\pi)^8}
                         \left[ {4603055\over 81} +{271360\over 27}\zeta_3
                            - 17600\zeta_5 \right],
   \end{array}
\end{eqnarray}
with $\zeta_3=1.20206\ldots$ and $\zeta_5= 1.03693\ldots$,
$\zeta$ being the Riemann zeta function.

We may immediately integrate eq.~(\ref{beta_def}) to obtain
\begin{equation}
   M = \Lambda^{\cal S} \left[b_0 g^{\cal S}(M)^2
                          \right]^{b_1\over 2b_0^2}
         \exp{\left[{1\over 2b_0 g^{\cal S}(M)^2}\right]} 
         \exp{\left\{ \int_0^{g^{\cal S}(M)} d\xi
          \left[ {1 \over \beta^{\cal S}(\xi)} +
                 {1\over b_0 \xi^3} - {b_1\over b_0^2\xi} \right]\right\} }.
\label{lambda_def}
\end{equation}
The renormalisation group invariant ({\it RGI})
quark mass is defined from the renormalised quark mass as
\begin{equation}
   m^{RGI} \equiv \Delta Z_m^{\cal S}(M) m^{\cal S}(M)
               = \Delta Z_m^{\cal S}(M) Z_m^{\cal S}(M) m_{bare},
\label{mrgi_msbar}
\end{equation}
where
\begin{equation}
   [\Delta Z_m^{\cal S}(M)]^{-1} = 
          \left[ 2b_0 g^{\cal S}(M)^2 \right]^{d_{m0}\over 2b_0}
          \exp{\left\{ \int_0^{g^{\cal S}(M)} d\xi
          \left[ {\gamma_m^{\cal S}(\xi)
                             \over \beta^{\cal S}(\xi)} +
                 {d_{m0}\over b_0 \xi} \right] \right\} },
\label{deltam_def}
\end{equation}
and the integration constant upon integrating eq.~(\ref{beta_def})
is given by $\Lambda^{\cal S}$, and similarly from eq.~(\ref{gamma_def})
we have $m^{RGI}$. $\Lambda^{\cal S}$ and $m^{RGI}$
are independent of the scale. Under a change of variable
(scheme change or ${\cal S}\to {\cal S}^\prime$),
\begin{equation}
   g^{{\cal S}^\prime} = G(g^{\cal S}) = 
               g^{\cal S}(1 + c_1 (g^{\cal S})^2 + \ldots).
\label{G_def}
\end{equation}
It can be shown that the first two coefficients of the $\beta$ function,
the first coefficient of the $\gamma$ function and $m^{RGI}$ are
independent of the scheme, while $\Lambda$ only changes as
$\Lambda^{{\cal S}^\prime} = \Lambda^{\cal S} \exp(c_1/b_0)$.

\begin{figure}[t]
   \hspace*{2.00cm}
   \epsfxsize=10.00cm \epsfbox{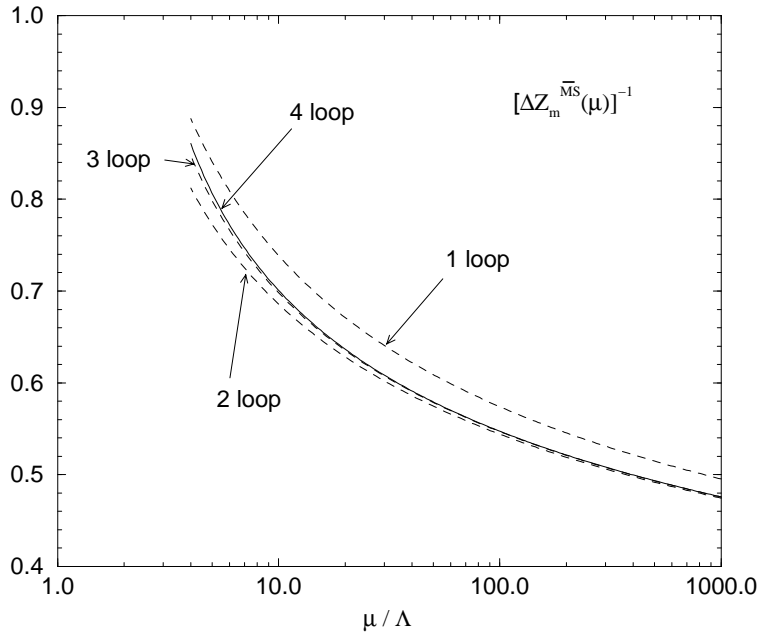}
   \caption{\it One-, two-, three- and four-loop results for
            $[\Delta Z_m^{\msbar}(\mu)]^{-1}$ in units of
            $\Lambda^{\msbar}$.}
   \label{fig_mMsbar_o_mrgi_lat99}
\end{figure}

For the $\overline{MS}$ scheme computing
$[\Delta Z_m^{\msbar}(\mu)]^{-1}$ involves first solving
eq.~(\ref{lambda_def}) for $g^{\msbar}(\mu)$ and then evaluating
eq.~(\ref{deltam_def}). This gives Fig.~\ref{fig_mMsbar_o_mrgi_lat99}.
We expand the $\beta$ and $\gamma$ functions to the appropriate
order and then numerically evaluate the integrals.
At $\mu = 2\,\mbox{GeV}$ we have $\mu / \Lambda^{\msbar} \sim 8$,
and it seems that already at this value we have a fast
converging series in loop orders.
Indeed, only going from one loop to two loops gives a significant change in
$[\Delta Z_m^{\msbar}(8\Lambda^{\msbar})]^{-1}$ of order $7\%$.
From two loops to three loops we have about $2\%$. The difference
between the three-loop and four-loop results is $O(0.5\%)$.
So if we are given $m^{RGI}$, and we wish to find the quark mass in the
$\overline{MS}$ scheme at a certain scale, we need only use the four-loop
result from eq.~(\ref{deltam_def}) or equivalently
Fig.~\ref{fig_mMsbar_o_mrgi_lat99}.

% ----------------------------------------------------------------------

\section{Digression: which scale to use?}
\label{scale_digression}

\begin{figure}[t]
   \hspace*{2.00cm}
   \epsfxsize=10.00cm \epsfbox{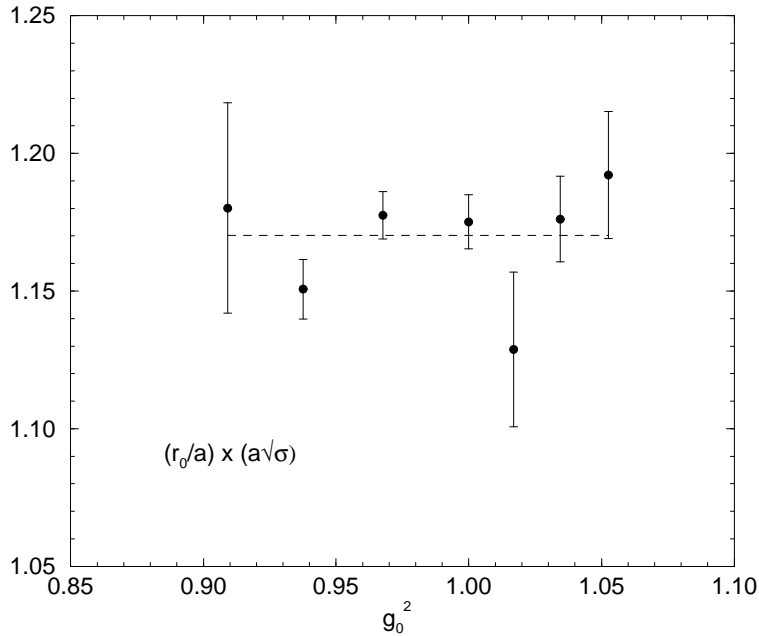}
   \caption{\it The product of $r_0/a$ and the string tension
            $a\sqrt{\sigma}$. $(r_0/a)(g_0)$ is taken from the
            formula given in \cite{guagnelli98a}, while the string tension 
            is taken from \cite{bali93a} ($\beta = $ $5.7$, $5.8$,
            $5.9$, $6.4$), \cite{bali97a} ($\beta = $ $6.0$, $6.2$),
            \cite{bali93b} ($\beta = 6.4$), \cite{bali98a} 
            ($\beta = 6.6$).}
   \label{fig_string_b+r_r0sqrtK_lat99}
\end{figure}

We always need one (or more) experimental numbers as input to set the
scale. Ideally, it should not matter what quantity we use. Obvious
choices are the force scale $r_0$ \cite{sommer94a}, or
the string tension $\sqrt{\sigma}$, or some particle
mass (e.g. the proton, or for quenched QCD at least, the $\rho$).
So a first requirement is that whatever quantity we use, we should be in
a region where the scaling to the continuum limit is the same for all
quantities. Thus for $r_0$ and the string tension we wish that
\begin{equation}
   { r_0 \over a}(g_0)
          \times (a\sqrt{\sigma})(g_0)  = \mbox{const}.
\label{product_r0_st}
\end{equation}
over the $g_0^2 \equiv 6/\beta$ region used in the simulations,
and indeed for all smaller $g_0^2$.
(We know that this must break down below a value of
$\beta$ around $5.7$,
due to the appearance of non-universal terms.)
In Fig.~\ref{fig_string_b+r_r0sqrtK_lat99} we show this product.
This seems reasonably constant, with a fit value of $1.170(5)$.

The second requirement is to set the scale in $\mbox{MeV}$.
As we are considering quenched 
QCD, it is not obvious that choosing scales from different
experimental (or phenomenological) quantities will necessarily
lead to the same results. Indeed, in the real world
\cite{sommer94a,gockeler97a} the values are
\begin{eqnarray}
   r_0 &=& 0.5\,\mbox{fm} \equiv (394.6\,\mbox{MeV})^{-1},
                                              \nonumber \\
   \sqrt{\sigma} &=& 427\,\mbox{MeV},
\label{different_scales}
\end{eqnarray}
($1\,\mbox{fm}^{-1} = 197.3\,\mbox{MeV}$)
which gives for the product a value of $1.082$ -- almost a $10\%$ difference
from the quenched value. As both phenomenological estimates come
from the same potential model \cite{eichten80a},
presumably the quenched lattice potential
has a slightly different shape from the continuum potential. 

Recently, the ALPHA collaboration has determined a value
for $\Lambda^{\msbar}$ \cite{capitani98a} of
\begin{equation}
   \Lambda^{\msbar} = 0.602(48) / r_0,
\label{lambda_res}
\end{equation}
which may easily be converted using eq.~(\ref{product_r0_st})
to the string tension scale. However, the numerical value
will then suffer from the same $10\%$ ambiguity. Thus we find
\begin{eqnarray}
   \Lambda^{\msbar}_{r_0} &=&  238(19)\,\mbox{MeV},
                                         \nonumber    \\ 
   \Lambda^{\msbar}_{\sqrt{\sigma}}
                                 &=&  220(18)\,\mbox{MeV}.
\end{eqnarray}
Our recent spectrum results \cite{pleiter99a}, using $O(a)$
improved fermions, also show a difference
whether one uses the $\rho$ or the proton mass to set the scale.
Using $r_0$ to set the scale
is roughly equivalent to using $m_\rho$.

In the following we shall adopt the $r_0$ scale as given in
\cite{guagnelli98a}, namely
\begin{equation}
   \ln (a / r_0) = -1.6805 - 1.7139(\beta-6) + 0.8155(\beta-6)^2
                   - 0.6667(\beta-6)^3
\label{r0_result}
\end{equation}
(with an error of $0.3\%$ increasing to $0.6\%$ for $\beta$ in the
range $5.7 \le \beta \le 6.57$), but delay using
a numerical value for this scale for as long as possible.
For the standard scale of $\mu = 2\,\mbox{GeV}$
this gives, upon solving eqs.~(\ref{lambda_def}) and (\ref{deltam_def})
to the appropriate loop order, the results for
$[\Delta Z_m^{\msbar}(\mu)]^{-1}$ shown in
Table~\ref{table_msbar_values}.
For later reference the results for some other $\mu$ values are also
given there, together with $\alpha_s^{\msbar}(\mu)$.

\begin{table}[htbp]
   \begin{center}
      \begin{tabular}{||l||l|l||l|l||}
         \hline
         \multicolumn{1}{||c||}{$\mu$}
                        & one-loop & two-loop & three-loop & four-loop  \\
         \hline
         \hline
         \multicolumn{1}{||c||}{} &
           \multicolumn{4}{c||}{} \\[-0.9em]
         \multicolumn{1}{||c||}{} &
           \multicolumn{4}{c||}{$[\Delta Z_m^{\msbar}(\mu)]^{-1}$} \\
         \multicolumn{1}{||c||}{} &
           \multicolumn{4}{c||}{} \\[-0.9em]
         \hline
         \hline
 $2.00\,\mbox{GeV}$            & $0.760(10)$ & $0.704(9)$
                               & $0.718(10)$ & $0.721(10)$  \\
 $2.12\,\mbox{GeV}$ ($1/a$ at $\beta = 6.0$)
                               & $0.752(10)$ & $0.697(8)$
                               & $0.711(9)$  & $0.714(10)$  \\
 $2.90\,\mbox{GeV}$ ($1/a$ at $\beta = 6.2$)
                               & $0.716(8)$  & $0.667(7)$
                               & $0.677(7)$  & $0.679(8)$   \\
 $3.85\,\mbox{GeV}$ ($1/a$ at $\beta = 6.4$)
                               & $0.689(7)$  & $0.644(6)$
                               & $0.652(5)$  & $0.653(6)$   \\
         \hline
         \hline
         \multicolumn{1}{||c||}{} &
           \multicolumn{4}{c||}{} \\[-0.9em]
         \multicolumn{1}{||c||}{} &
           \multicolumn{4}{c||}{$\alpha_s^{\msbar}(\mu)$} \\
         \multicolumn{1}{||c||}{} &
           \multicolumn{4}{c||}{} \\[-0.9em]
         \hline
 $2.00\,\mbox{GeV}$            & $0.268(10)$ & $0.195(6)$
                               & $0.201(6)$  & $0.202(7)$  \\
 $2.12\,\mbox{GeV}$ ($1/a$ at $\beta = 6.0$)
                               & $0.261(9)$  & $0.191(5)$
                               & $0.196(6)$  & $0.197(6)$  \\
 $2.90\,\mbox{GeV}$ ($1/a$ at $\beta = 6.2$)
                               & $0.228(7)$  & $0.170(5)$
                               & $0.174(5)$  & $0.175(5)$  \\
 $3.85\,\mbox{GeV}$ ($1/a$ at $\beta = 6.4$)
                               & $0.205(6)$  & $0.156(3)$
                               & $0.159(4)$  & $0.159(4)$ \\
         \hline
      \end{tabular}
   \end{center}
\caption{\it Useful values of $[\Delta Z_m^{\msbar}(\mu)]^{-1}$ and
         $\alpha_s^{\msbar}(\mu)
         \equiv (g^{\msbar}(\mu))^2/4\pi$. The errors are 
         a reflection of the error in eq.~(\ref{lambda_res}).
         The values of $1/a$ are found from eq.~(\ref{r0_result})
         together with $r_0$ from eq.~(\ref{different_scales}).}
\label{table_msbar_values}
\end{table}

% ----------------------------------------------------------------------

\section{Determining the quark mass}
\label{qm_determination}

We shall now derive formulae for the quark masses 
using the conserved vector current ({\it CVC}) and
the partially conserved conserved axial vector current ({\it PCAC})
by assuming Taylor expansions in the bare quark
mass for the relevant functions that occur. We distinguish two quark 
masses. The Ward identities arising from an infinitesimal vector
transformation in the partition function lead to a bare quark
mass given by
\begin{equation}
   am_{q_i} \equiv {1\over 2} \left( {1\over \kappa_{q_i}} 
                                   - {1 \over \kappa_c} \right),
                              \qquad i = 1,2,
\end{equation}
where $\kappa_{q_i}$ is the corresponding hopping parameter, and 
$\kappa_c$ is the critical hopping parameter. This is the standard 
definition of the quark mass.
Similarly, for an infinitesimal axial transformation
the Ward Identity ({\it WI}) or {\it PCAC}
definition of the quark mass can be written as
\begin{eqnarray}
   a\widetilde{m}_{q_1} + a\widetilde{m}_{q_2}
    &\stackrel{t\gg 0}{=}&
          { \langle
            \partial_4 {\cal A}_4^{q_1q_2}(t){\cal P}^{q_1q_2;smeared}(0)
          \rangle \over
            \langle {\cal P}^{q_1q_2}(t){\cal P}^{q_1q_2;smeared}(0) \rangle}
                                                    \label{pcac_corr} \\
    &\equiv&
       - \sinh am_{PS}^{q_1q_2}
            { \langle 0|
                        \hat{\cal A}_4^{q_1q_2} |PS \rangle \over
              \langle 0| \hat{\cal P}^{q_1q_2}   |PS \rangle },
                                                   \label{correlation_mq}
\end{eqnarray}
where ${\cal A}$ (${\cal P}$) is the axial vector current
(pseudoscalar density). The precise form of ${\cal A}$ and
and ${\cal P}$ will be given later for the $O(a)$ improved as well as
the Wilson cases. We have summed the operators over their spatial planes.
While ${\cal A}(t)$ and ${\cal P}(t)$ should be point operators, to improve 
the signal ${\cal P}(0)$ is smeared over its spatial plane.
To obtain the second equation, we have re-written eq.~(\ref{pcac_corr})
in a Fock space and then introduced a complete set of states in the usual
way. We have then picked out the lowest pseudoscalar ($PS$)
or $0^{-+}$ state whose mass we denote by $m_{PS}^{q_1q_2}$.

Both definitions of the quark mass must be renormalised.
In a scheme ${\cal S}$ at scale $M$ we have
\begin{eqnarray}
   m_{q_i}^{\cal S}(M) &=& Z_m^{\cal S}(M,am_{q_i}) m_{q_i},
                                                        \nonumber  \\
   m_{q_1}^{\cal S}(M) + m_{q_2}^{\cal S}(M)
                       &=& \widetilde{Z}_m^{\cal S}(M,am_{q_1},am_{q_2})
                           \left( \widetilde{m}_{q_1} + \widetilde{m}_{q_2}
                           \right).
\label{qm_renormalisation}
\end{eqnarray}
The Ward identities give $Z_m = 1/Z_S$ (from $CVC$) and
$\widetilde{Z}_m = Z_A/Z_P$ (from $PCAC$).

Let us now Taylor expand $\widetilde{m}$ and
the pseudoscalar mass $m_{PS}^{q_1q_2}$ in terms of the
bare quark masses $m_{q_i}$,
\begin{eqnarray}
   \lefteqn{\half ( a\widetilde{m}_{q_1} + a\widetilde{m}_{q_2}) =}
       &                                               \nonumber  \\
       &  \widetilde{Y} \left[ 1 + \widetilde{c}\half (am_{q_1} + am_{q_2})
                         + \widetilde{d} { (am_{q_1})^2 + (am_{q_2})^2 \over
                                 am_{q_1} + am_{q_2} } + \ldots
                  \right] \half (am_{q_1} + am_{q_2}),
                                                        \nonumber  \\
   \lefteqn{(am_{PS}^{q_1q_2})^2 =}
       &                                               \nonumber  \\
       &  Y_{PS} \left[ 1 + c_{PS}\half (am_{q_1} + am_{q_2})
                           + d_{PS} { (am_{q_1})^2 + (am_{q_2})^2 \over
                                 am_{q_1} + am_{q_2} } + \ldots
                  \right] \half (am_{q_1} + am_{q_2}).
\label{taylor_expansions}
\end{eqnarray}
The functions must be symmetric under interchange of the quarks, i.e.\
$q_1 \leftrightarrow q_2$.
Only at the lowest (first) order in the quark mass 
is the functional form simply $am_{q_1} + am_{q_2}$. At the next order both
terms, $(am_{q_1} + am_{q_2})^2$ and $(am_{q_1})^2 + (am_{q_2})^2$
are allowed. Taylor expanding eq.~(\ref{qm_renormalisation})
and comparing with eq.~(\ref{taylor_expansions}) gives us
\begin{equation}
   \widetilde{Y} = { Z^{\cal S}_m(M) \over \widetilde{Z}_m^{\cal S}(M)}
   \equiv { Z^{\cal S}_P(M) \over Z_S^{\cal S}(M)Z_A}.
\label{twiddles_Y}
\end{equation}
Renormalisation constants which show no explicit quark mass dependence refer 
to $m_{q_1} = m_{q_2} = 0$.

We shall also Taylor-expand the matrix elements appearing in
eq.~(\ref{correlation_mq}).
First we define the bare pseudoscalar decay constant by
$\langle 0|{\cal A}^{q_1q_2}_4 |PS \rangle = m^{q_1q_2}_{PS}
                                                    f^{q_1q_2}_{PS}$,
and similarly we set
$\langle 0|{\cal P}^{q_1q_2} |PS \rangle = - g^{q_1q_2}_{PS}$.
Expanding the decay constants $f^{q_1,q_2}_{PS}$ and $g^{q_1q_2}_{PS}$ 
to first order in the quark masses gives
\begin{equation}
   \widetilde{d} = d_{PS} \equiv d, 
\end{equation}
and hence
\begin{equation}
   {\half\left(a\widetilde{m}_{q_1} + a\widetilde{m}_{q_2}\right)
            \over (am_{PS}^{q_1q_2})^2}
        = {\widetilde{Y}\over Y_{PS}}
          \left[ 1 + \left( { \widetilde{c} - c_{PS} \over Y_{PS} } \right)
                     (am_{PS}^{q_1q_2})^2 + \ldots 
          \right].
\label{practical_end_result}
\end{equation}
Thus, at least to this order, we have a relation between the
{\it WI} quark masses and the pseudoscalar mass.

Using eq.~(\ref{qm_renormalisation}) gives the renormalised quark mass,
and we additionally use eq.~(\ref{mrgi_msbar}) to re-write this in
a {\it RGI} form as
\begin{equation}
   {\half (r_0 m^{RGI}_{q_1} + r_0 m^{RGI}_{q_2}) \over
        (r_0 m^{q_1q_2}_{PS})^2}
       = c^*_a + c^*_b (r_0 m^{q_1q_2}_{PS})^2 + \ldots ,
\label{end_result}
\end{equation}
with 
\begin{equation}
   c^*_a = \lim_{g_0\to 0} c_a(g_0), \qquad
   c^*_b = \lim_{g_0\to 0} c_b(g_0),
\label{c_values_lim}
\end{equation}
and
\begin{eqnarray}
   c_a &=& \left[ \Delta Z_m^{\cal S}(M) \widetilde{Z}_m^{\cal S}(M) \right]
                  \times \left[{\widetilde{Y}\over Y_{PS}}\right] \times
                 \left({r_0\over a}\right)^{-1},
                                                        \nonumber \\
   c_b &=& \left[ \Delta Z_m^{\cal S}(M) \widetilde{Z}_m^{\cal S}(M) \right]
                  \times \left[{\widetilde{Y}\over Y_{PS}}\right]
                  \times \left[{-c_{PS}\over Y_{PS}}
                               \left({r_0\over a}\right)^{-2}\right]
                  \times \left({r_0\over a}\right)^{-1}.
\label{ca+cb_defs}
\end{eqnarray}
Upon taking the continuum limit $g_0 \to 0$, any scaling violations
will show themselves as non-constant $c_a$, $c_b$ functions.
Equation~(\ref{end_result}) is the main result of this analysis.
Given $c^*_a$, $c^*_b$ and the pseudoscalar mass,
we can then determine the quark masses.

For the $K^+$ ($u\bar{s}$) we set $q_1=u$, $q_2=s$, and
for the $K^0$ ($d\bar{s}$) we set $q_1=d$, $q_2=s$.
Together with the $\pi^+$ ($u\bar{d}$), with $q_1=u$, $q_2=d$,
this gives from eq.~(\ref{end_result})
\begin{eqnarray}
   r_0 m^{RGI}_s &=& c^*_a \left[ (r_0 m_{K^+})^2 + (r_0 m_{K^0})^2
                                    - (r_0 m_{\pi^+})^2 \right] +
                                           \nonumber \\
                 & & \qquad \qquad
                     c^*_b \left[ (r_0 m_{K^+})^4 + (r_0 m_{K^0})^4
                                    - (r_0 m_{\pi^+})^4 \right] + \ldots,
                                           \nonumber \\
   r_0 m^{RGI}_l &=& c^*_a (r_0 m_{\pi^+})^2 +
                     c^*_b (r_0 m_{\pi^+})^4 + \ldots,
\label{m_rgi}
\end{eqnarray}
where we have defined $m^{RGI}_l = (m^{RGI}_u + m^{RGI}_d)/2$, i.e.\ the
average of the $u/d$ quarks. We have ignored any small corrections
due to electromagnetic effects.

% ----------------------------------------------------------------------

\section{Numerical results}
\label{numerical_results}

% ----------------------------------------------------------------------

\subsection{Pseudoscalar mesons and bare quark masses}

For degenerate quark masses from eq.~(\ref{taylor_expansions}) we have
\begin{eqnarray}
   a\widetilde{m}_q &=& \widetilde{Y}
                    \left[ 1 + (\widetilde{c}+d)am_q + \ldots \right] am_q,
                                                        \nonumber  \\
   (am_{PS})^2  &=& Y_{PS} \left[ 1 + (c_{PS}+d)am_q + \ldots \right] am_q,
\label{qm_degen}
\end{eqnarray}
and
\begin{equation}
   { a\widetilde{m}_q \over (am_{PS})^2 } = {\widetilde{Y}\over Y_{PS}}
                \left[ 1 + \left( { \widetilde{c} - c_{PS} \over Y_{PS} }
                           \right) (am_{PS})^2 + \ldots 
                \right],
\label{ps_degen}
\end{equation}
where $m_{PS} \equiv m_{PS}^{qq}$ (i.e.\ $q_1=q_2 \equiv q$).
Equation~(\ref{ps_degen}) gives $\widetilde{Y}/Y_{PS}$
for the $c_a$ term in eq.~(\ref{m_rgi}), but the gradient 
$(\widetilde{c}-c_{PS})/ Y_{PS}$
is not sufficient to give $-c_{PS}/ Y_{PS}$ for the $c_b$ term. 
For $O(a)$ improved fermions, associating the mass expansion parameters
$b_A$, $b_P$ and $b_m$ \cite{luscher96b} with our
expansions, we find $\widetilde{c} \equiv -(b_A-b_P)$
and $d \equiv b_m$. First order perturbation theory
\cite{sint96a} gives $\widetilde{c} \sim 0.001g_0^2$.~\footnote{A 
non-perturbative estimate \cite{divitiis97a}, however,
gives $\widetilde{c} \sim -0.15$ at $\beta=6.2$.} On top of that 
$\widetilde{c}/c_{PS} = O(a)$, so that the effect of $\widetilde{c}$ can 
safely be ignored. For Wilson fermions we shall 
assume that either $\widetilde{c}$ is small in comparison
with $c_{PS}$, as above, or that the complete term $c_b (r_0m_K)^2$
is small when compared with $c_a$. As we shall see, little error is 
introduced by this assumption. 

We have computed the pseudoscalar mass $m_{PS}$ and the {\it WI}
bare quark mass both, for $O(a)$ improved fermions and Wilson fermions. 
For improved fermions the calculations were done at $\beta = 6.0, 6.2, 6.4$
and $c_{SW} = 1.769, 1.614, 1.526$ \cite{luscher96b}, respectively, while 
for Wilson fermions we only did calculations at $\beta = 6.0, 6.2$. 
The computational methods used are standard.
For the pseudoscalar mass we used the correlation function
\begin{eqnarray}
   C(t) &=& \langle {\cal P}^{smeared}(t){\cal P}^{smeared}(0) \rangle
                                         \nonumber \\
        &\stackrel{t\gg 0}{=}& A \left[ e^{-m_{PS}t} + e^{-m_{PS}(T-t)}
                                \right],
\end{eqnarray}
$T$ being the temporal extent of the lattice.
To improve the signal, a Jacobi-smeared operator was used,
as described in \cite{gockeler97a}. For Wilson fermions
the pseudoscalar meson mass results can also be found in \cite{gockeler97a}.
The $O(a)$ improved results for $m_{PS}$
are given in Table~\ref{table_run_params} in the Appendix.
The various $\kappa$ values used,
the lattice size, and the number of configurations generated
are also collated there.

For $a\widetilde{m}_q$ the ratio of two point correlation functions
as given in eq.~(\ref{pcac_corr}) was used.
For $O(a)$ improved fermions, as well as the action,
the operators must also be improved:
\begin{eqnarray}
   {\cal A}_\mu
          &=& A_\mu + c_A a\partial_\mu P,
                                           \nonumber \\
   {\cal P}
          &=& P,
\end{eqnarray}
where $A_\mu = \overline{q}\gamma_\mu\gamma_5 q$ and
$P = \overline{q}\gamma_5 q$. By choosing the improvement
coefficient $c_A(g_0)$ appropriately, the Ward identity
can be made exact to $O(a)$. $c_A(g_0)$ is non-perturbatively known
\cite{luscher96a}. In Table~\ref{table_wiqm} in the Appendix
we give our results for $a\widetilde{m}_q$.

\begin{figure}[t]
   \vspace*{-0.10in}
   \hspace*{2.00cm}
   \epsfxsize=10.00cm \epsfbox{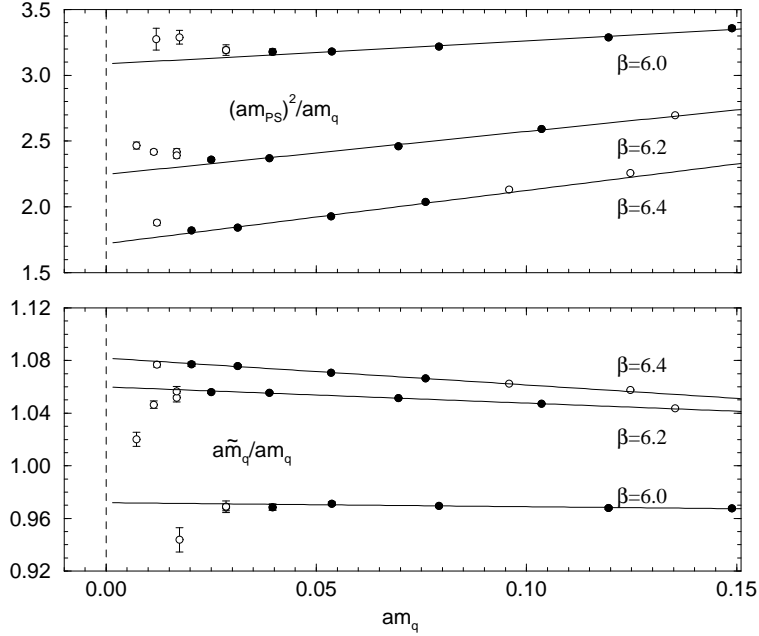}
   \caption{\it $(am_{PS})^2/am_q$ and $a\widetilde{m}_q/am_q$ against
            $am_q$ for $O(a)$ improved fermions.
            Filled circles denote points used in the fits.}
   \label{fig_ratio_magic_lat99}
\end{figure}

\begin{figure}[t]
   \hspace*{2.00cm}
   \epsfxsize=10.00cm \epsfbox{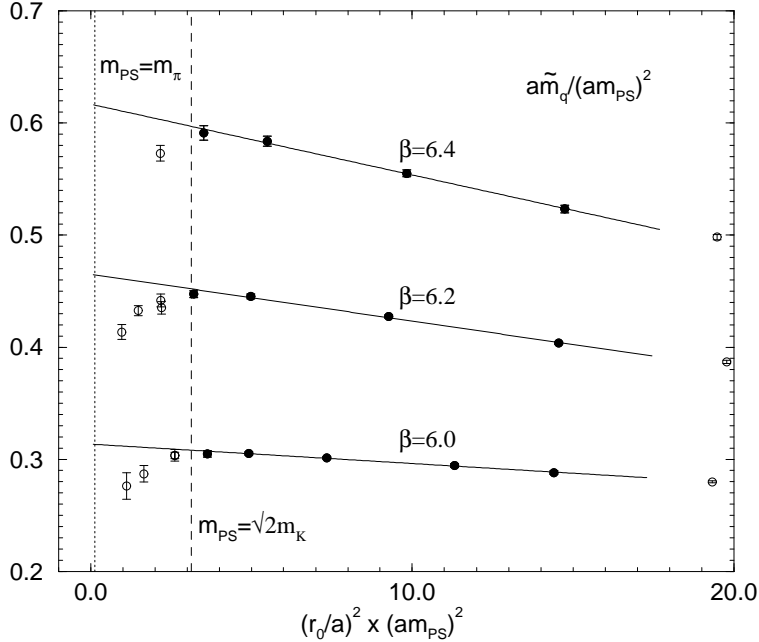}
   \caption{\it $a\widetilde{m}_q/(am_{PS})^2$ against
   $((r_0/a)\times(am_{PS}))^2$
   for $O(a)$ improved fermions. Filled circles denote points used in the 
   fits. The dashed line ($\sim 3.13$)
   is $m_{PS} = \sqrt{2} m_K$ (which here corresponds to a 
   fictitious $s\overline{s}$ bound state) while the dotted line
   ($\sim 0.125$) is $m_\pi$.}
   \label{fig_mqwiompi2_mpi2_magic_lat99}
\end{figure}

Let us first discuss $O(a)$ improved fermions.
In Fig.~\ref{fig_ratio_magic_lat99}
we show the ratios $(am_{PS})^2/am_q$ and $a\widetilde{m}_q/am_q$
against $am_q$, while in Fig.~\ref{fig_mqwiompi2_mpi2_magic_lat99}
we plot $a\widetilde{m}_q/(am_{PS})^2$ against
$((r_0/a)\times(am_{PS}))^2$.
We must now search for a region where eq.~(\ref{qm_degen}),
without higher order terms, is valid. For large quark mass values
we expect non-linear terms, while for small quark masses
quenched QCD chiral logarithms become significant.
Finite volume effects do not seem to be a problem, as for
$\beta = 6.0$, $\kappa =0.1342$ and $\beta = 6.2$, $\kappa = 0.1352$ 
we have made runs on two different volumes, without significant
changes in the results.

We now make some cuts. In Fig.~\ref{fig_mqwiompi2_mpi2_magic_lat99}
we see that for small quark masses there are significant deviations
from linearity. In particular the light quark mass $m_l$ lies
in a region where no direct linear extrapolation is possible.
However, above $m_{PS} \approx \sqrt{2} m_K$ deviations from linearity
seem small. We shall thus assume that at least above the
strange quark mass any effects of chiral logarithms
are small. For heavy quark masses, on the other hand,
linearity is still present until at least $m_q \approx 3m_s \approx 
\third m_c$.
(Note that $2(r_0 m_D)^2 \sim 44.9$). In this interval
lie four or more quark masses. The results of the various fits
are given in Table~\ref{table_qm_fits}.
\begin{table}[htbp]
   \begin{center}
      \begin{tabular}{||l|l|l|l||}
         \hline
 \multicolumn{1}{||c}{}&
 \multicolumn{1}{|c}{} &
 \multicolumn{1}{|c}
          {} &
 \multicolumn{1}{|c||}{}                      \\[-0.9em]
 \multicolumn{1}{||c}{$\beta$}&
 \multicolumn{1}{|c}{${\widetilde{Y}\over Y_{PS}}$} &
 \multicolumn{1}{|c}
          {${\widetilde{Y}\over Y_{PS}}\times
        {\widetilde{c}-c_{PS}\over Y_{PS}} \left({r_0\over a}\right)^{-2}$}&
 \multicolumn{1}{|c||}{$\widetilde{Y}$}             \\[-0.9em]
 \multicolumn{1}{||c}{}&
 \multicolumn{1}{|c}{} &
 \multicolumn{1}{|c}
          {} &
 \multicolumn{1}{|c||}{}                      \\
         \hline
         \multicolumn{4}{||c||}{$O(a)$ improved fermions}       \\
         \hline
 6.0      & 0.314(2)     & -0.0017(2) & 0.972(6)   \\
 6.2      & 0.465(2)     & -0.0042(2) & 1.060(3)   \\
 6.4      & 0.617(5)     & -0.0063(5) & 1.082(5)   \\
         \hline
         \multicolumn{4}{||c||}{Wilson fermions}                \\
         \hline
 6.0      & 0.368(4)   & -0.00025(51)& 0.711(12)   \\
 6.2      & 0.502(6)   & -0.0036(5)  & 0.814(10)   \\
         \hline
      \end{tabular}
   \end{center}
\caption{\it Fit values.}
\label{table_qm_fits}
\end{table}

\begin{figure}[tp]
   \hspace*{2.00cm}
   \epsfxsize=10.00cm \epsfbox{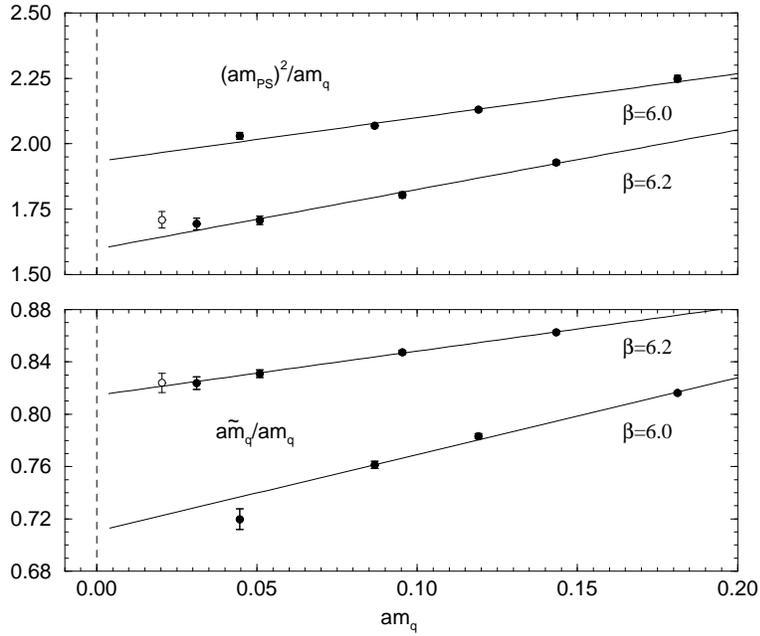}
   \caption{\it $(am_{PS})^2/am_q$ and $a\widetilde{m}_q/am_q$ against
            $am_q$ for Wilson fermions.}
   \label{fig_ratio_wilson_lat99}
\end{figure}

\begin{figure}[tp]
   \hspace*{2.00cm}
   \epsfxsize=10.00cm \epsfbox{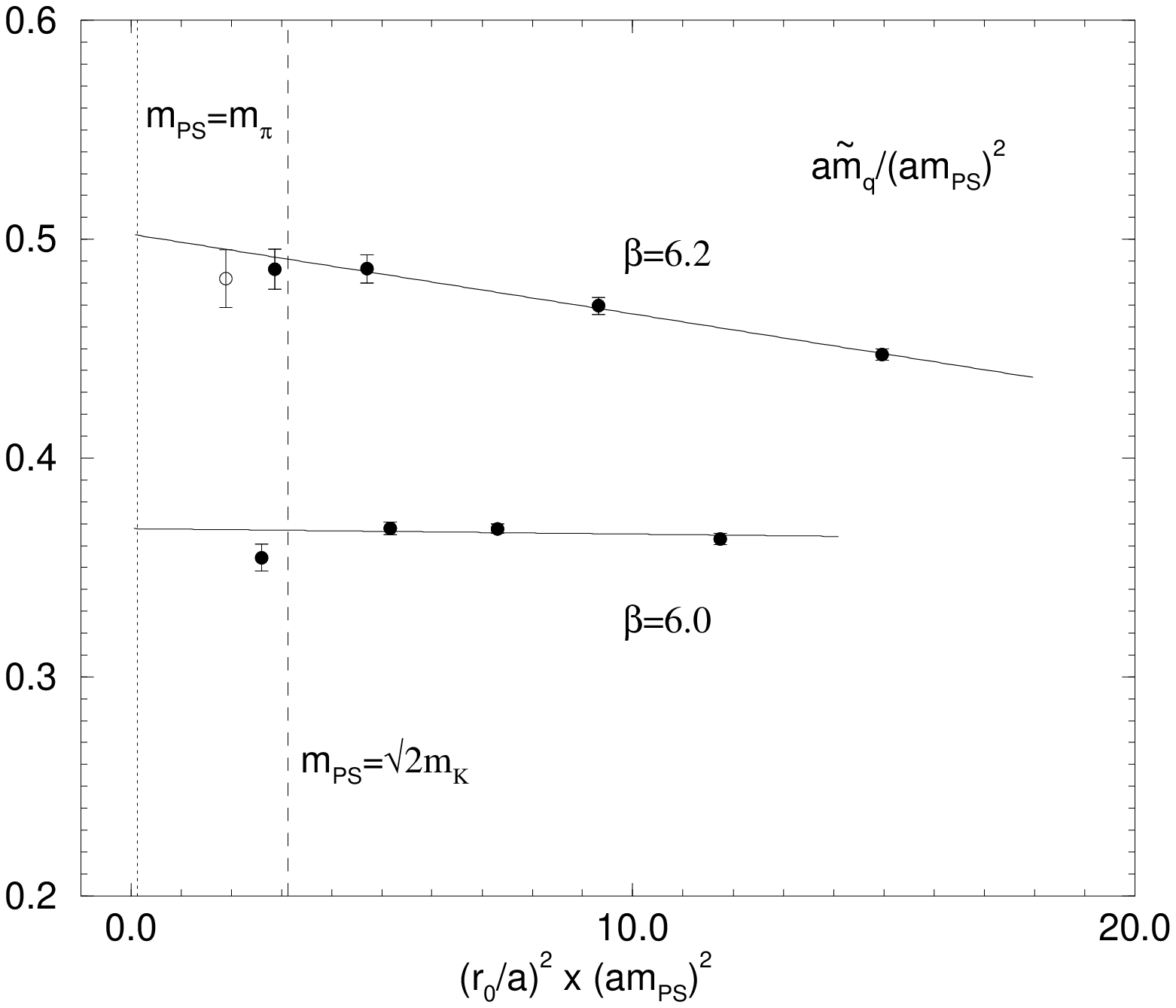}
   \caption{\it The same as Fig.~\ref{fig_mqwiompi2_mpi2_magic_lat99}
   but for Wilson fermions.}
   \label{fig_mqwiompi2_mpi2_wilson_lat99}
\end{figure}

For Wilson fermions we simply set ${\cal A}_\mu = A_\mu$.
The results for $a\widetilde{m}_q$ have also been given in
\cite{gockeler97a}. 
In Figs.~\ref{fig_ratio_wilson_lat99} and
\ref{fig_mqwiompi2_mpi2_wilson_lat99}
we show the ratios $(am_{PS})^2/am_q$, $a\widetilde{m}_q/am_q$ and
$a\widetilde{m}_q/(am_{PS})^2$.
Similar fit ranges as for improved fermions seem appropriate once more.

\begin{figure}[t]
   \hspace*{2.00cm}
   \epsfxsize=10.00cm \epsfbox{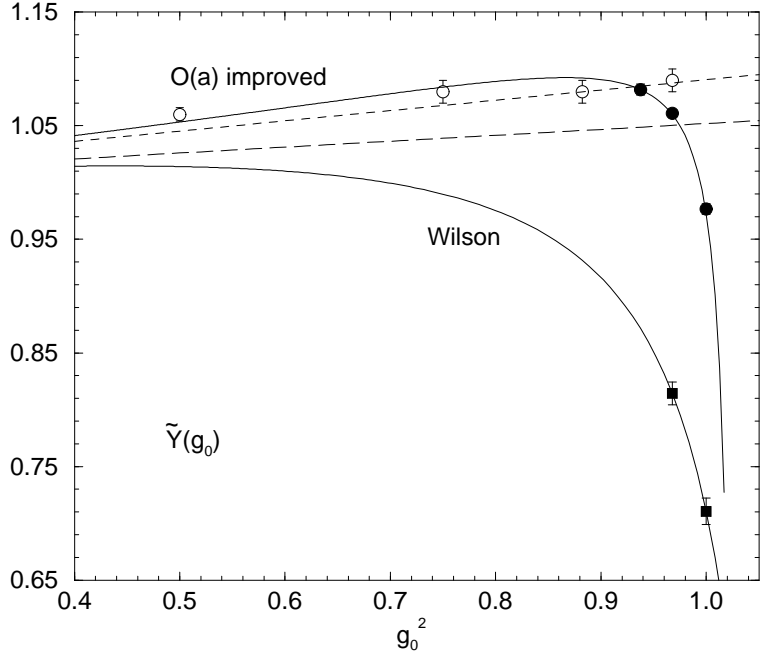}
   \caption{\it $\widetilde{Y}$ against $g_0^2$.
            Our $O(a)$ improved fermion results are shown as
            filled circles, while those from \cite{divitiis97a}
            are shown as open circles. The Wilson fermion results
            are filled squares.
            The one-loop perturbation theory results are
            also shown for the $O(a)$ improved case (dashed line)
            and the Wilson case (long dashed line).}
   \label{fig_ZmxZpoZa_magic+wilson_lat99}
\end{figure}

To illustrate the $g_0^2$ dependence of some of these results, we show in
Fig.~\ref{fig_ZmxZpoZa_magic+wilson_lat99}
the results for $\widetilde{Y}$ taken from
Table~\ref{table_qm_fits} together with Pad{\'{e}}-like interpolations
of the form
\begin{equation}
   \widetilde{Y}(g_0) = { 1 + p_1 g_0^2 + p_2 g_0^4 \over 
                      1 + (p_1-c)g_0^2 + p_3 g_0^4 },
\label{pade}
\end{equation}   
arranged so that the perturbative result
$\widetilde{Y}(g_0) = 1 + c g_0^2$ with $c = 0.09051$ for
improved fermions and $0.05195$ for Wilson fermions
is obtained for small $g_0^2$.
Possible Pad{\'{e}} interpolations are found to be $(p_1,p_2,p_3)=$
$(-1.24, 0.256, 0.347)$ for $O(a)$ improved fermions
and $(-0.944,0.00,0.0746)$ for Wilson fermions.
Also shown for comparison are $O(a)$ improved results found
in \cite{divitiis97a}. We see that for $O(a)$ improved fermions
first order perturbation theory is good for $g_0^2 \lsim 0.96$,
while for Wilson fermions a breakdown occurs much earlier.

Thus we now have estimates for
$\widetilde{Y}/Y_{PS}$ and $(\widetilde{c}-c_{PS})/Y_{PS}$.
$\widetilde{Y}$ will also be needed for Wilson fermions.

% ----------------------------------------------------------------------

\subsection{Renormalisation}

To compute $c_a$ and $c_b$ we must now determine the
factor $\Delta Z^{\cal S}_m(M) Z^{\cal S}_m(M)$.
For $O(a)$ improved fermions this was done by the ALPHA collaboration
\cite{capitani98a} using the Schr\"odinger Functional ({\it SF}) method.
With the notation of eq.~(\ref{mrgi_msbar}) their result can be written as
\begin{equation}
   \Delta Z_m^{SF}(L^{-1}) Z_m^{SF}(L^{-1}) = 1.752 + 0.321(\beta -6)
                                              - 0.220(\beta -6)^2
\label{F_fun}
\end{equation}
(valid for $6.0 \le \beta \le 6.5$),
where they have worked at a scale given in terms
of the box size $L$. As emphasised previously, this is a mapping
from the bare quark mass to the {\it RGI} mass, so this function is the same
in all schemes. In eq.~(\ref{F_fun}) the total error is about $2\%$.

\begin{figure}[tp]
   \hspace*{0.875in}
   \epsfxsize=10.00cm \epsfbox{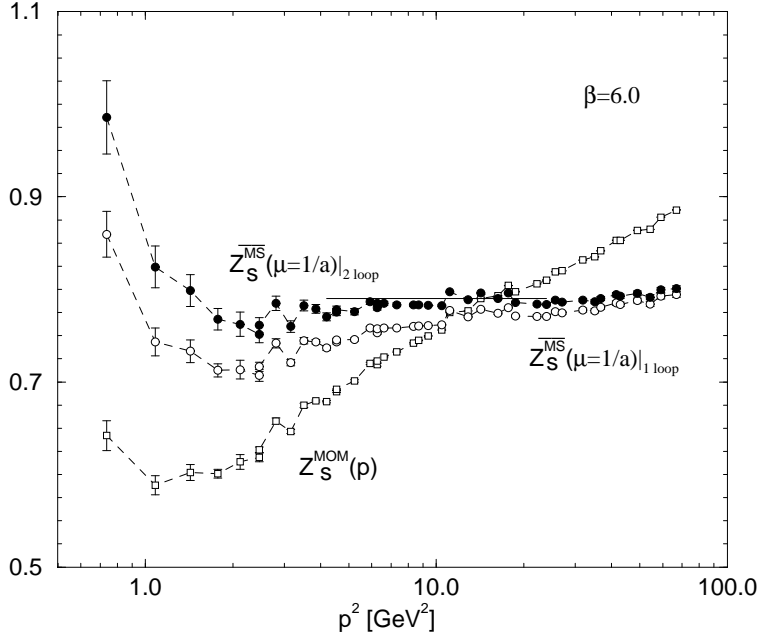}
   \caption{\it $Z^{\msbar}_S(\mu =1/a)$ against $p^2$ for
            $\beta=6.0$. (See \cite{gockeler98a} for details.)
            The open squares are the original data,
            in the chiral limit, while the filled circles 
            represent the results of multiplying $Z^{\mom}(p)$ by $X$
            using both coefficients $B_1$ and $B_2$.
            The open circles are the result of using only
            the $B_1$ coefficient. The straight line is a
            fit to the plateau, the length denoting the fit range chosen.}
   \label{fig_Zs_p2.b6p00kc_lat99}
\end{figure}

For Wilson fermions we use the method proposed in \cite{martinelli94a}
and refined in \cite{gockeler98a}.
This mimics perturbation theory in a certain {\it MOM} scheme
by considering amputated quark Green's functions in, say, the Landau
gauge, with an appropriate operator insertion.
The renormalisation constant is fixed at some scale $p^2$.
This gives a non-perturbative determination
of $Z_S^{\mom}(p)$. ($Z_P^{\mom}(p)$ is not suitable,
as chiral symmetry breaking means that $Z^{\mom}_P \to 0$ as 
we approach the chiral limit, as recently emphasised in \cite{cudell98a}.)
More details of the method, our momentum source approach, and results
are given in \cite{gockeler98a}.
As $\Delta Z^{\msbar}_m$ is known in the $\overline{MS}$
scheme at scale $\mu$ (see Fig.~\ref{fig_mMsbar_o_mrgi_lat99}
and Table~\ref{table_msbar_values}), we can write
\begin{equation}
   \Delta Z_m^{\mom}(p) \widetilde{Z}_m^{\mom}(p) 
      = { \Delta Z_m^{\msbar}(\mu) \over Z_S^{\msbar}(\mu)
                                                \widetilde{Y} }
      \equiv F(\beta),
\label{Delta_Z_wilson}
\end{equation}
where we have used eq.~(\ref{twiddles_Y}) and the definitions $Z_m =1/Z_S$,
$\Delta Z_m =1/\Delta Z_S$ and
\begin{equation}
   Z_S^{\msbar}(\mu) = 
        X_{S;\mom}^{\phantom{S;}\msbar}(\mu,p) Z_S^{\msbar}(p),
\label{convert}
\end{equation}
with
\begin{equation} 
   X_{S;\mom}^{\phantom{S;}\msbar}(\mu,p)
      = { \Delta Z_S^{\mom}(p) \over \Delta Z_S^{\msbar}(\mu) }.
\end{equation}
Here $X$ converts the renormalisation constant from the {\it MOM} scheme
to the $\overline{MS}$ scheme and can be calculated using
a continuum regularisation (e.g. naive dimensional regularisation).
So we can write
\begin{eqnarray}
   X_{S;\mom}^{\phantom{S;}\msbar}(\mu,\mu)
                   &=&  {Z_S^{\msbar}(\mu) \over
                         Z_S^{\mom}(\mu) }
                                          \nonumber   \\
                   &=&
        1 + {\alpha_s^{\msbar}(\mu) \over 4\pi} B_1
          + \left({\alpha_s^{\msbar}(\mu)
                              \over 4\pi}\right)^2 B_2
          + \ldots,
\end{eqnarray}
with $B_1 = 16/3$ \cite{gockeler97a}.
The coefficient $B_2$ has recently been calculated \cite{franco98a},
giving in the Landau gauge for $N_f = 0$ flavours a value
of $177.48452$. Hence, knowing the coefficients $B_1$, $B_2$,
we can trace back to the more general expression for $X$.
A suitable formula is
\begin{eqnarray}
   \lefteqn{[X_{S;\mom}^{\phantom{S;}\msbar}(\mu,p)]^{-1} =}
     &  &                                                \nonumber \\
     &  &
     \exp{ \left[ \int^{g^{\msbar}(\mu)}_{g^{\msbar}(p)} d\xi
                  {\gamma_m^{\msbar}(\xi)
                                      \over \beta^{\msbar}(\xi)} +
                  \int^{g^{\msbar}(p)}_0 d\xi
                  \left( {\gamma_m^{\msbar}(\xi) -
                          \gamma_m^{\mom}(G(\xi)) \over
                         \beta^{\msbar}(\xi)}
                  \right)
           \right] },
\end{eqnarray}
($g^{\mom}=G(g^{\msbar})$ from eq.~(\ref{G_def})).
Armed with this estimate for $X$, then from eq.~(\ref{convert})
we see that $Z^{\msbar}_S(\mu)$ should be independent of $p^2$, 
so plotting $Z^{\msbar}_S(\mu)$ against $p^2$ we expect to see
a plateau. In Fig.~\ref{fig_Zs_p2.b6p00kc_lat99} we show this,
plotting first the original data, extrapolated to the chiral limit,
then the results for $Z^{\msbar}_S$
when only using $B_1$, and finally using both $B_1$ and $B_2$.
We see the data for $Z_S$ becoming flatter.
The results of the fit are given in
the first column of Table~\ref{table_wilson_Z_fits}.
The appropriate values for $\widetilde{Y}$
(see Table~\ref{table_qm_fits}) and $\Delta Z^{\msbar}_m(\mu=1/a)$
(see Table~\ref{table_msbar_values}) are substituted in
eq.~(\ref{Delta_Z_wilson}) to give the results listed in
the second column of this table.
\begin{table}[htbp]
   \begin{center}
      \begin{tabular}{||c|c|c||}
         \hline
  &   & \\[-0.9em]
 $\beta$  &  $Z_S^{\msbar}(\mu=1/a)$ 
          &  $\Delta Z_m^{\mom}(p) \widetilde{Z}_m^{\mom}(p)$ \\
  &   & \\[-0.9em]
         \hline
 6.0      & 0.790(5) & 2.49(6) \\
 6.2      & 0.803(5) & 2.25(4) \\
         \hline
      \end{tabular}
   \end{center}
\caption{\it Fit values for the Wilson renormalisation constants.}
\label{table_wilson_Z_fits}
\end{table}

% ----------------------------------------------------------------------

\section{Continuum results}
\label{results}

\begin{figure}[tp]
   \begin{tabular}{cc}
      \hspace{-0.50cm}
      \epsfxsize=7.00cm \epsfbox{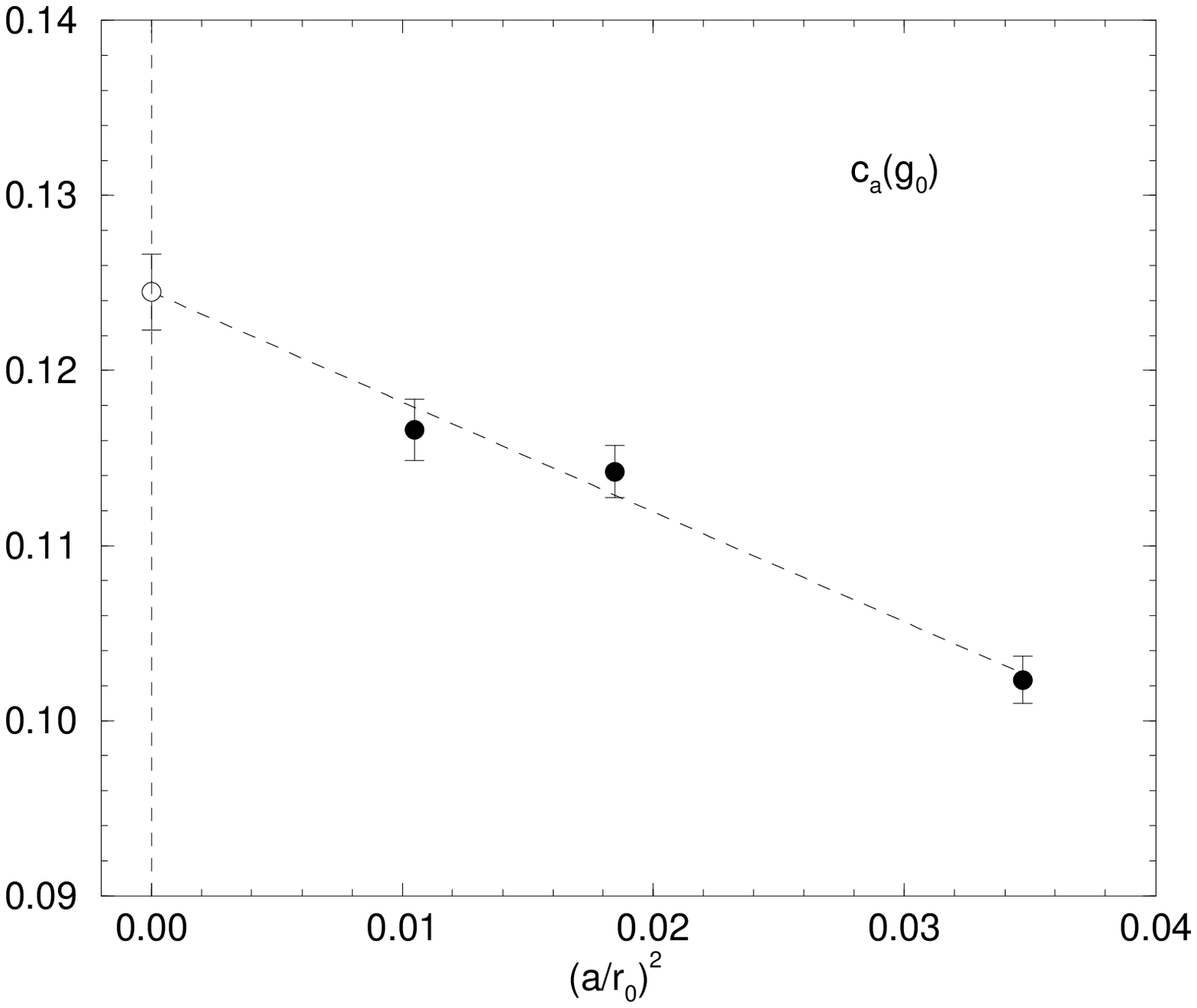} &
      \epsfxsize=7.50cm \epsfbox{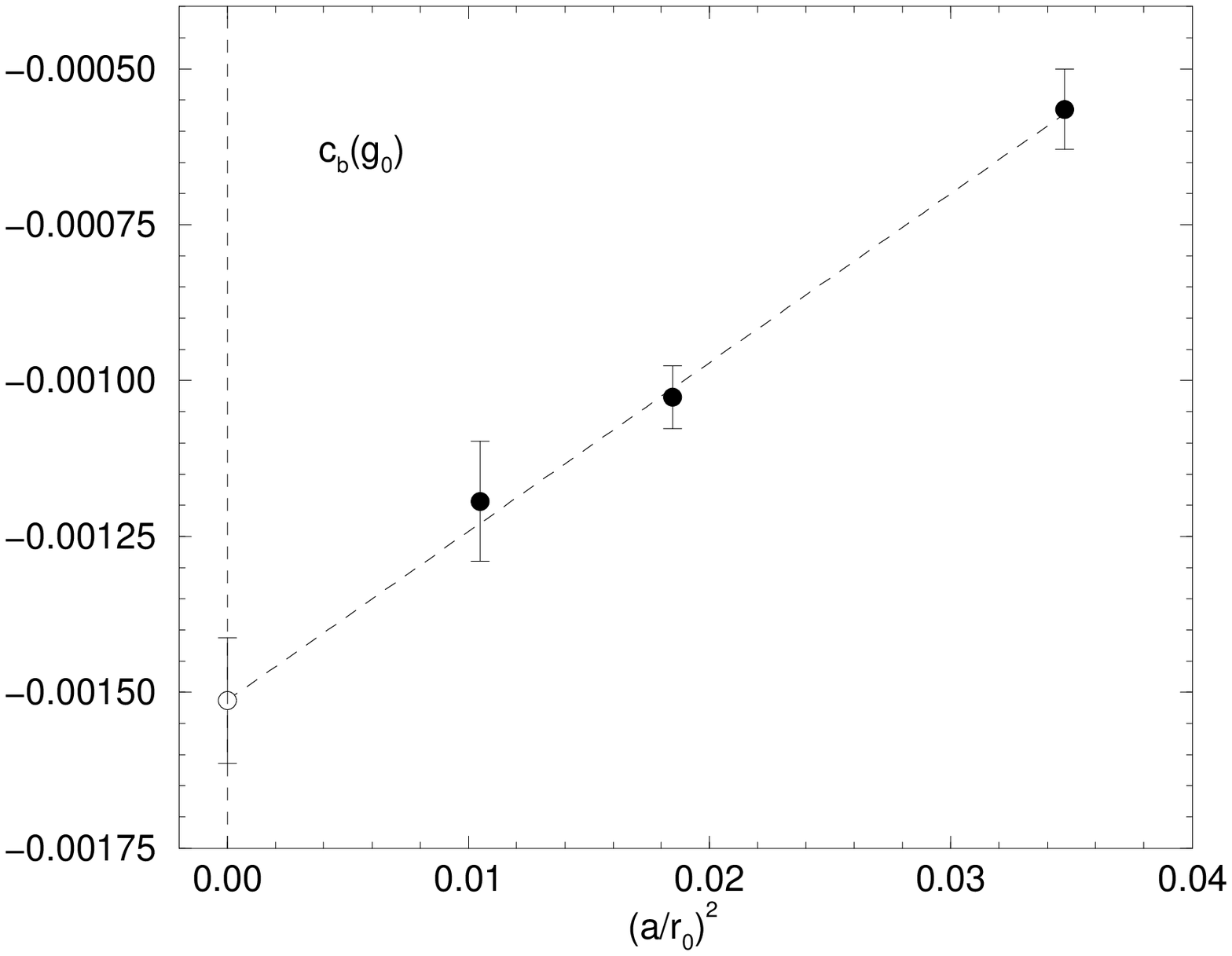}
   \end{tabular}
\caption{\it The continuum extrapolation for $c_a$ and $c_b$ for
         $O(a)$ improved fermions.}
\label{c_extrapolation_magic}
\end{figure}

\begin{figure}[tp]
   \begin{tabular}{cc}
      \hspace{-0.50cm}
      \epsfxsize=7.00cm \epsfbox{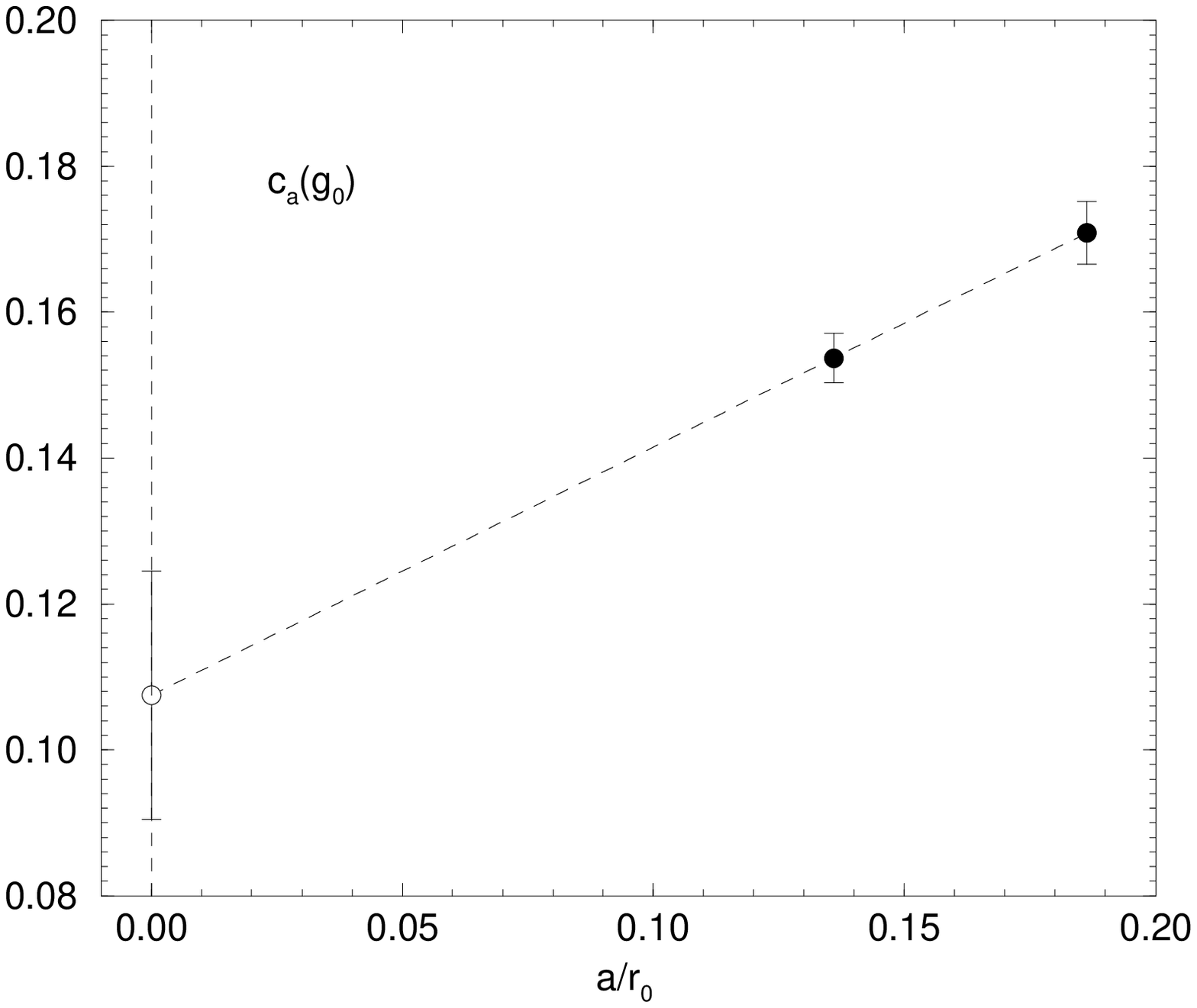} &
      \epsfxsize=7.25cm \epsfbox{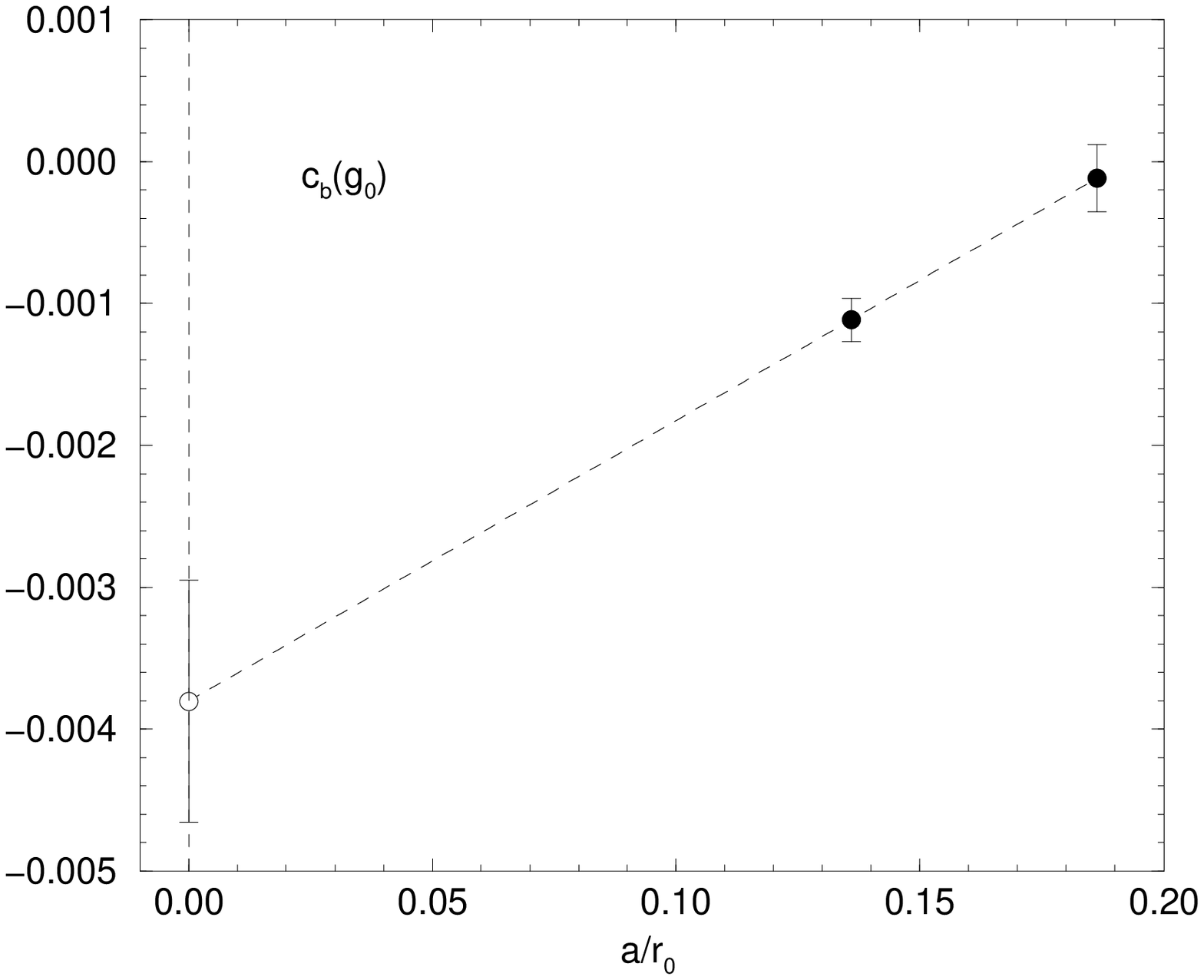}
   \end{tabular}
\caption{\it The continuum extrapolation for $c_a$ and $c_b$ for
         Wilson fermions.}
\label{c_extrapolation_wilson}
\end{figure}

Plotting $c_a$ and $c_b$ against $a^2$ for $O(a)$ improved
fermions, and against $a$ for Wilson fermions, 
we can now extrapolate these coefficients to the continuum limit.
In Fig.~\ref{c_extrapolation_magic}
we show the results for $c_a$ and $c_b$ for improved fermions.
A linear fit is also plotted. The results of this fit are given
in Table~\ref{table_c_fits}.
As anticipated, using the first order perturbative result
from \cite{sint96a} for $\widetilde{c}$ in $c_b$ has
no influence on the result.~\footnote{Note, however, that the 
non-perturbative estimate from
\cite{divitiis97a} at $\beta = 6.2$ is
$\widetilde{c}/Y_{PS}\times (r_0/a)^{-2} \sim -0.0012$ and has a similar
order of magnitude to the kept term.}
In Fig.~\ref{c_extrapolation_wilson} we show the equivalent results
for Wilson fermions.
In this case, as we only have two $\beta$ values,
the fit degenerates to an extrapolation. The results of this
extrapolation are also given in Table~\ref{table_c_fits}.
We note that the results for $O(a)$ improved fermions
and Wilson fermions for $c_a^*$ are compatible with each other.
\begin{table}[htbp]
    \begin{center}
        \begin{tabular}{||c|l|l||}
           \hline
           \multicolumn{1}{||c}{}            &
           \multicolumn{1}{|c}{$c_a^*$}      &
           \multicolumn{1}{|c||}{$c_b^*$}    \\
           \hline
              $O(a)$ improved & 0.124(2)   & -0.0015(1)      \\
              Wilson          & 0.107(17)  & -0.0038(9)      \\
           \hline
        \end{tabular}
    \end{center}
\caption{\it The continuum extrapolation of $c_a \to c_a^*$
         and $c_b \to c_b^*$ for
         $O(a)$ improved fermions and Wilson fermions.}
\label{table_c_fits}
\end{table}

Upon inserting these numbers in eq.~(\ref{m_rgi}),
we find our estimate for the {\it RGI} strange and $u/d$
quark masses. To convert to physical numbers,
we now have in quenched QCD the
uncertainty in the scale, as discussed in section ~\ref{scale_digression}.
If we use the scale $r_0=0.5\,\mbox{fm}$, eq.~(\ref{different_scales}),
then together with the experimental values of the $\pi$ and $K$ masses,
namely $m_{\pi^+} = 139.6\,\mbox{MeV}$ and $m_{K^+} = 493.7\,\mbox{MeV}$,
$m_{K^0} = 497.7\,\mbox{MeV}$, we find the results
for $O(a)$ improved fermions
\begin{eqnarray}
   m_s^{RGI} &=& 146(4)\,\mbox{MeV},
                                           \nonumber \\
   m_l^{RGI} &=& 6.1(2)\,\,\mbox{MeV}.
\end{eqnarray}
The error comes from Table~\ref{table_c_fits} and from 
eq.~(\ref{F_fun}). As can also be seen from Table~\ref{table_c_fits},
most of the result comes from the constant term, with the slope
giving only a small correction to the answer.

For Wilson fermions we have $m_s^{RGI} = 121(20)\,\mbox{MeV}$,
$m_l^{RGI} = 5.3(8)\,\mbox{MeV}$.
This result is somewhat lower than the $O(a)$ improved
numbers. We ascribe this mainly to the fact that we
only have two values of $\beta$, which makes a continuum
extrapolation more difficult. Also the number of $\kappa$ values
used and the size of the data sets are smaller than
for $O(a)$ improved fermions. Nevertheless, within a one-standard
deviation the results are in agreement.

In the $\overline{MS}$ scheme at the `standard' value
of $\mu = 2\,\mbox{GeV}$, using the four-loop results from
Table~\ref{table_msbar_values}, we find for $O(a)$ improved fermions
\begin{eqnarray}
    m^{\msbar}_s(\mu=2\,\mbox{GeV}) &=& 105(4)\,\mbox{MeV},
                                           \nonumber \\
    m^{\msbar}_l(\mu=2\,\mbox{GeV}) &=& 4.4(2)\,\,\mbox{MeV}.
\label{mq_msbar}
\end{eqnarray}
The corresponding Wilson results are 
$m^{\msbar}_s(\mu=2\,\mbox{GeV}) = 87(15)\,\mbox{MeV}$ and
$m^{\msbar}_l(\mu=2\,\mbox{GeV}) = 3.8(6)\,\mbox{MeV}$.

Note that for the $m_l$ quark mass result we have simply
extrapolated the fits for the strange quark mass result downwards.
The mass ratio $m_s/m_l$ for $O(a)$ improved fermions
is $\approx 23.9$, which is very close to the value given in
leading order chiral perturbation theory -- namely
$(m_{K^+}^2 + m_{K^0}^2-m_{\pi^+}^2) / m_{\pi^+}^2 \approx 24.2$
(see eq.~(\ref{m_rgi})).
This is simply because $|c_b^*| \ll |c_a^*|/(r_0m_K)^2$, and so
the second term in eq.~(\ref{m_rgi}) is almost negligible. The mass
ratio is then independent of $c_a^*$.

\begin{figure}[t]
   \hspace*{0.875in}
   \epsfxsize=10.00cm \epsfbox{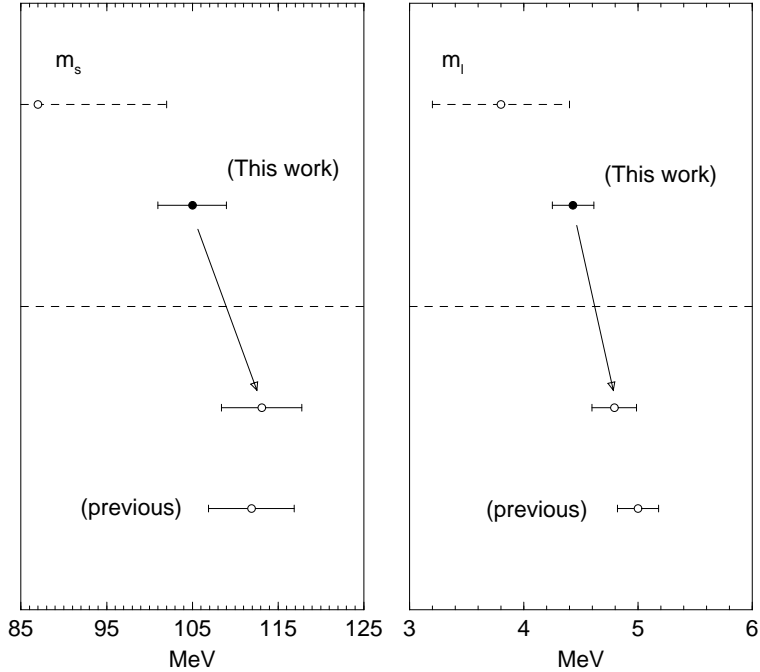}
   \caption{\it The strange and light quark masses
            from eq.~(\ref{mq_msbar}) (`This work'). 
            The Wilson fermion results are shown dotted.
            Also shown is our previous result
            (`previous') \cite{gockeler97a}. The arrows
            denote the result when using $\sqrt{\sigma}$ as a scale.}
\label{fig_qm_comparison}
\end{figure}

In Fig.~\ref{fig_qm_comparison} we plot our results.
Below the dotted line, we have given our previous result \cite{gockeler97a},
using the string tension as the scale, as given in 
eq.~(\ref{different_scales}).
As a comparison we have also re-plotted our result given in
eq.~(\ref{mq_msbar}) using $\sqrt{\sigma}$ as the scale.
A reasonable agreement is seen. As our previous result used
tadpole improved ({\it TI}) perturbation theory to
determine the renormalisation constant, it would seem that the use
of {\it TI} perturbation theory does not introduce much error.
In the next section we shall briefly investigate this point.

% ----------------------------------------------------------------------

\section{Digression: comparison with tadpole-improv\-ed perturbation theory}
\label{TI}

In this section we shall discuss how reliable {\it TI}
perturbation theory is. Lowest order perturbation theory gives
\begin{equation}
   Z_S^{\msbar}(\mu = 1/a)
                 = 1 - {g_0^2 \over 16\pi^2 }C_F
                    B^{\msbar}(c_{SW}) + O(g_0^4).
\label{perturbative_result}
\end{equation}
The $B$ coefficient is uncomfortably large \cite{gockeler97a}.
In \cite{lepage92a} this was traced to large tadpole
diagrams in the perturbation expansion and also to the
expansion in a non-physical (bare) coupling constant.
Removing the tapole diagrams and expanding in (say)
$\alpha_s^{\msbar}$ gives the improved series
\begin{equation}
   Z_S^{\msbar}(\mu = 1/a)
                 = u_0 \left( 
                      1 - {\alpha_s^{\msbar}(\mu=1/a) \over 4\pi} C_F
                          \left[B^{\msbar}(\widetilde{c}_{SW})
                                - \pi^2 \right] \right)  + O(\alpha_s^2),
\label{TI_perturbative_result}
\end{equation}
with $u_0 = \langle \third\mbox{Tr} U_{plaq} \rangle^{\quarter}$
to be numerically determined, and $\widetilde{c}_{SW}=c_{SW}u_0^3$.
A description of our variation of this procedure is given
in \cite{gockeler97a}. In particular, we choose $c_{SW}$ to be the
non-perturbatively determined result, rather than setting it to be
equal to $1/u_0^3$. $\alpha_s^{\overline{MS}}$ corresponds
to the four-loop results (see Table~\ref{table_msbar_values})
also used in \cite{lepage92a}.

As we now have a genuine non-perturbative determination of $Z_m$
available, it is of interest to compare how the different results scale
to the continuum limit. It is convenient to first define for
two $O(a)$ improved estimations of the renormalisation constant the ratio
\begin{equation}
   R = {\Delta Z^{{\cal S}_1}_m(M_1) Z_m^{{\cal S}_1}(M_1) \over
        \Delta Z^{{\cal S}_2}_m(M_2) Z_m^{{\cal S}_2}(M_2)}
     = 1 + O(a^2).
\label{ratio_definition}
\end{equation}
For ${\cal S}_2$ we choose the {\it SF} scheme and the result given
in eq.~(\ref{F_fun}), while for ${\cal S}_1$ we use the $\overline{MS}$
scheme, together with either the perturbative result,
eq.~(\ref{perturbative_result}), or the $TI$ result,
eq.~(\ref{TI_perturbative_result}).

\begin{figure}[t]
   \hspace*{0.750in}
   \epsfxsize=10.00cm \epsfbox{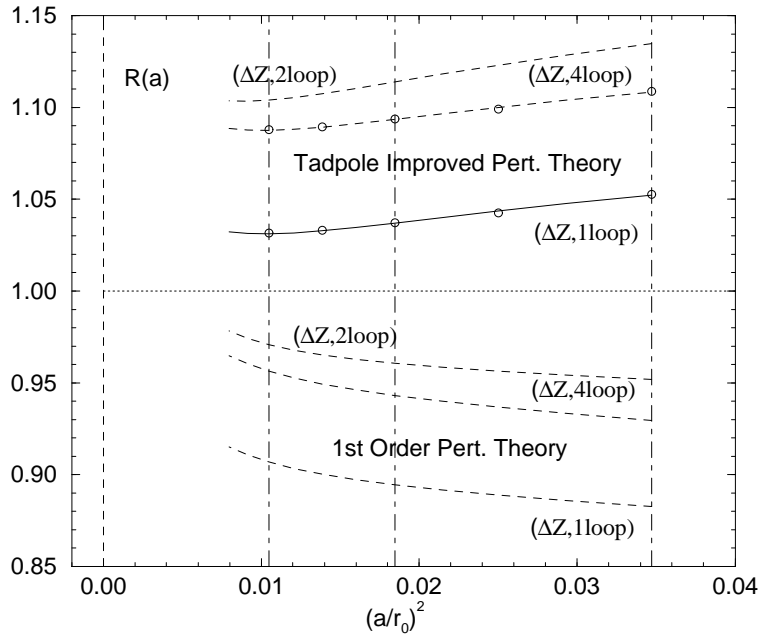}
   \caption{\it The results for the ratio $R$ as defined in
            eq.~(\ref{ratio_definition}). $Z_m^{\overline{MS}}$
            has been determined from eq.~(\ref{perturbative_result})
            or eq.~(\ref{TI_perturbative_result})
            and $\Delta Z_m^{\msbar}$ taken from
            Table~\ref{table_msbar_values}. A simple interpolation 
            has been used between the $\beta$ values.
            The dot-dashed vertical lines correspond to 
            $\beta = 6.0$, $6.2$, $6.4$, from right to left.} 
\label{fig_zm_rgi_a2_magic.R_lat99}
\end{figure}

In Fig.~\ref{fig_zm_rgi_a2_magic.R_lat99} we show the results
for the ratio $R$
for perturbation theory and {\it TI} perturbation theory, using
consistently the one-, two- and four-loop results from
Table~\ref{table_msbar_values}. As we originally used
one-loop perturbation theory results, it seems more
consistent to also use the one-loop result for $\Delta Z_m$
when converting $Z^{\overline{MS}}_S$ to the $RGI$ result. 
This gives the solid line in the figure. This was the approach
adopted in \cite{gockeler97a}.
We expect $O(a)$ effects to become apparent as $a^2\to 0$
if ${\cal S}_1$ is not exactly $O(a)$ improved. However,
a linear fit in $a^2$ for the {\it TI} result (with one-loop $\Delta Z_m$)
appears to go to $R=1$ with an error of only about $2\%$,
while for the equivalent perturbative result
the error is about $10\%$. Thus, in this case tadpole improving the 
perturbative result does give better results.
However, choosing other loop orders changes the picture somewhat
and can make using perturbation theory a better choice. 
We would like to emphasise that this picture does not
have to hold for other renormalisation constants. Strictly speaking a 
case-by-case analysis is required.

% ----------------------------------------------------------------------

\section{Conclusions}
\label{conclusions}

In this article we have calculated the strange and $u/d$ quark
masses for quenched QCD, both for $O(a)$ improved fermions
and Wilson fermions, using a non-per\-turb\-atively determined
renormalisation constant. Our results are given in eq.~(\ref{mq_msbar}) 
and the lines that follow it. 
Corrections to leading order chiral perturbation theory
are small, if we stay away from the region where chiral logarithms become
significant. We have also seen that using {\it TI} perturbation theory
rather than simple perturbation theory does not automatically lead
to an improvement of the continuum result.

% ----------------------------------------------------------------------

\section*{Acknowledgements}
\label{acknowledgement}

The numerical calculations were performed on the
Quadrics {\it QH2} at DESY (Zeut\-hen) as well as
the Cray {\it T3E} at ZIB (Berlin) and the Cray {\it T3E} at
NIC (J\"ulich). We wish to thank all institutions for their support.

% ----------------------------------------------------------------------

\section*{Note added}

While this work was being completed, we received a
copy of \cite{garden99a}. This contains some similar results to ours.

% ----------------------------------------------------------------------

\newpage

\appendix

\section*{Appendix}

In Table~\ref{table_run_params} we give our parameter values
used in the $O(a)$ improved fermion simulations together with
the measured pseudoscalar mass.
For most of the overlapping values with \cite{gockeler97a}
there has been some increase in statistics.
\begin{table}[th]
\begin{small}
   \begin{center}
      \begin{tabular}{||l|l|l|l|c||l||}
         \hline
\multicolumn{1}{||c}{$\beta$} &
\multicolumn{1}{|c}{$c_{sw}$} &
\multicolumn{1}{|c}{$\kappa$} & 
\multicolumn{1}{|c}{Volume} & \multicolumn{1}{|c||}{$\#$ configs.} 
& \multicolumn{1}{c||}{$am_{PS}$}  \\
         \hline
 6.0  & 1.769  &  0.1217   & $16^3\times 32$ & $O(150)$      & 1.2546(12)\\
 6.0  & 1.769  &  0.1263   & $16^3\times 32$ & $O(150)$      & 0.9704(11)\\
 6.0  & 1.769  &  0.1285   & $16^3\times 32$ & $O(150)$      & 0.8189(11)\\
 6.0  & 1.769  &  0.1300   & $16^3\times 32$ & $O(160)$      & 0.7071(16)\\
 6.0  & 1.769  &  0.1310   & $16^3\times 32$ & $O(160)$      & 0.6268(16)\\
 6.0  & 1.769  &  0.1324   & $16^3\times 32$ & $O(990)$      & 0.5042(7) \\
 6.0  & 1.769  &  0.1333   & $16^3\times 32$ & $O(990)$      & 0.4122(9) \\
 6.0  & 1.769  &  0.1338   & $16^3\times 32$ & $O(520)$      & 0.3549(12)\\
 6.0  & 1.769  &  0.1342   & $16^3\times 32$ & $O(1300)$     & 0.3012(10)\\
 6.0  & 1.769  &  0.1342   & $24^3\times 32$ & $O(200)$      & 0.3017(13)\\
 6.0  & 1.769  &  0.1346   & $24^3\times 32$ & $O(200)$      & 0.2390(12)\\
 6.0  & 1.769  &  0.1348   & $24^3\times 32$ & $O(200)$      & 0.1978(16)\\
         \hline
         \hline
 6.2  & 1.614  & 0.1247    & $24^3\times 48$ & $O(100)$      & 1.0284(9) \\
 6.2  & 1.614  & 0.1294    & $24^3\times 48$ & $O(100)$      & 0.7217(9) \\
 6.2  & 1.614  & 0.1310    & $24^3\times 48$ & $O(100)$      & 0.6043(9) \\
 6.2  & 1.614  & 0.1321    & $24^3\times 48$ & $O(260)$      & 0.5183(6) \\
 6.2  & 1.614  & 0.1333    & $24^3\times 48$ & $O(560)$      & 0.4136(6) \\
 6.2  & 1.614  & 0.1344    & $24^3\times 48$ & $O(560)$      & 0.3034(6) \\
 6.2  & 1.614  & 0.1349    & $24^3\times 48$ & $O(560)$      & 0.2431(7) \\
 6.2  & 1.614  & 0.1352    & $24^3\times 48$ & $O(260)$      & 0.2016(10)\\
 6.2  & 1.614  & 0.1352    & $32^3\times 64$ & $O(110)$      & 0.2005(9) \\
 6.2  & 1.614  & 0.1354    & $32^3\times 64$ & $O(290)$      & 0.1657(6) \\
 6.2  & 1.614  & 0.13555   & $32^3\times 64$ & $O(280)$      & 0.1339(7) \\
         \hline
         \hline
 6.4  & 1.526  & 0.1313    & $32^3\times 48$ & $O(100)$      & 0.5305(9) \\
 6.4  & 1.526  & 0.1323    & $32^3\times 48$ & $O(100)$      & 0.4522(10)\\
 6.4  & 1.526  & 0.1330    & $32^3\times 48$ & $O(100)$      & 0.3935(12)\\
 6.4  & 1.526  & 0.1338    & $32^3\times 48$ & $O(200)$      & 0.3213(8) \\
 6.4  & 1.526  & 0.1346    & $32^3\times 48$ & $O(200)$      & 0.2402(8) \\
 6.4  & 1.526  & 0.1350    & $32^3\times 48$ & $O(200)$      & 0.1923(9) \\
 6.4  & 1.526  & 0.1353    & $32^3\times 64$ & $O(260)$      & 0.1507(8) \\ 
         \hline
      \end{tabular}
   \end{center}
\caption{\it Parameter values used in the simulations, together with
             the measured pseudoscalar mass.}
\label{table_run_params}
\end{small}
\end{table}

The results for the {\it WI} quark mass, $a\widetilde{m}_q$, are
first split into two pieces. $2a\widetilde{m}^{(0)}_q$ denotes the
mass coming from the
$\langle \partial_4 A_4 P^{smeared} \rangle / 
\langle P P^{smeared}\rangle$
ratio, while $2a\widetilde{m}^{(1)}_q$ is the result of
$\langle \nabla^2_4 P P^{smeared} \rangle / \langle P P^{smeared}\rangle$.
The sum
$2a\widetilde{m}_q = 2a\widetilde{m}_q^{(0)} + 2c_A a\widetilde{m}_q^{(1)}$
gives the {\it WI} quark mass. All these results are given in
Table~\ref{table_wiqm}.
We define
$(\partial_4)_{xy} \equiv (\delta_{x+\hat{4},y}-\delta_{x-\hat{4},y})/2$.
$\partial_4\partial_4$ has been replaced by
$(\nabla^2_4)_{x,y} \equiv \delta_{x+\hat{4},y} - 2\delta_{x,y} +
                          \delta_{x-\hat{4},y}$.
In the continuum limit both $\partial_4\partial_4$ and
$\nabla^2_4$ give the same derivative. 
On the lattice we choose the discretisation 
$\nabla^2_4$ with the smallest (temporal) extension.
In \cite{gockeler97a} the choice $\partial_4\partial_4$ was used.
\begin{table}[th]
\begin{small}
   \begin{center}
      \begin{tabular}{||l|l|l|l|l||l||}
         \hline
\multicolumn{1}{||c}{$\beta$}  &
\multicolumn{1}{|c}{$c_{sw}$}  &
\multicolumn{1}{|c}{$\kappa$}  & 
\multicolumn{1}{|c}{$2a\widetilde{m}_q^{(0)}$} &
\multicolumn{1}{|c||}{$2a\widetilde{m}_q^{(1)}$} & 
\multicolumn{1}{c||}{$2a\widetilde{m}_q$} \\
         \hline
 6.0  & 1.769  & 0.1217    & 0.9677(7)       & 1.7867(33)    & 0.8197(5)  \\
 6.0  & 1.769  & 0.1263    & 0.5936(4)       & 1.0173(24)    & 0.5093(3)  \\
 6.0  & 1.769  & 0.1285    & 0.4340(3)       & 0.7080(22)    & 0.3754(3)  \\
 6.0  & 1.769  & 0.1300    & 0.3314(3)       & 0.5227(21)    & 0.2881(3)  \\
 6.0  & 1.769  & 0.1310    & 0.2651(3)       & 0.4072(20)    & 0.2313(3)  \\
 6.0  & 1.769  & 0.1324    & 0.1751(1)       & 0.2609(8)     & 0.1535(1)  \\
 6.0  & 1.769  & 0.1333    & 0.1186(1)       & 0.1737(7)     & 0.1042(1)  \\
 6.0  & 1.769  & 0.1338    & 0.08734(21)     & 0.1274(9)     & 0.07678(18)\\
 6.0  & 1.769  & 0.1342    & 0.06282(16)     & 0.09217(62)   & 0.05519(15)\\
 6.0  & 1.769  & 0.1342    & 0.06294(27)     & 0.0932(12)    & 0.05522(25)\\
 6.0  & 1.769  & 0.1346    & 0.03772(32)     & 0.0584(11)    & 0.03289(32)\\
 6.0  & 1.769  & 0.1348    & 0.02485(37)     & 0.0399(12)    & 0.02154(37)\\
         \hline
         \hline
 6.2  & 1.614  & 0.1247    & 0.7342(3)       & 1.1520(18)    & 0.6915(2)  \\
 6.2  & 1.614  & 0.1294    & 0.4011(2)       & 0.5451(12)    & 0.3809(2)  \\
 6.2  & 1.614  & 0.1310    & 0.2966(2)       & 0.3776(10)    & 0.2826(2)  \\
 6.2  & 1.614  & 0.1321    & 0.2272(1)       & 0.2748(7)     & 0.2170(1)  \\
 6.2  & 1.614  & 0.1333    & 0.15265(5)      & 0.1741(4)     & 0.14620(5) \\ 
 6.2  & 1.614  & 0.1344    & 0.08542(6)      & 0.09336(36)   & 0.08196(5) \\
 6.2  & 1.614  & 0.1349    & 0.05510(6)      & 0.05985(34)   & 0.05288(6) \\
 6.2  & 1.614  & 0.1352    & 0.03688(11)     & 0.04073(46)   & 0.03537(11)\\
 6.2  & 1.614  & 0.1352    & 0.03703(12)     & 0.04046(37)   & 0.03553(12)\\
 6.2  & 1.614  & 0.1354    & 0.02478(6)      & 0.02773(18)   & 0.02375(6) \\
 6.2  & 1.614  & 0.13555   & 0.01549(8)      & 0.01802(20)   & 0.01482(8) \\ 
         \hline
         \hline
 6.4  & 1.526  & 0.1313    & 0.2709(1)       & 0.2870(9)     & 0.2637(1)  \\
 6.4  & 1.526  & 0.1323    & 0.2090(1)       & 0.2074(9)     & 0.2038(1)  \\
 6.4  & 1.526  & 0.1330    & 0.1660(1)       & 0.1567(8)     & 0.1621(1)  \\
 6.4  & 1.526  & 0.1338    & 0.1172(1)       & 0.1045(4)     & 0.11461(6) \\
 6.4  & 1.526  & 0.1346    & 0.06882(6)      & 0.05799(38)   & 0.06736(6) \\
 6.4  & 1.526  & 0.1350    & 0.04467(7)      & 0.03739(35)   & 0.04373(6) \\ 
 6.4  & 1.526  & 0.1353    & 0.02661(5)      & 0.02276(22)   & 0.02603(5) \\
         \hline
      \end{tabular}
   \end{center}
\caption{\it Results for the Ward Identity quark mass $2a\widetilde{m}_q$.}
\label{table_wiqm}
\end{small}
\end{table}

\clearpage

% ----------------------------------------------------------------------

% ----------------------------------------------------------------------

%
\end{document}